\documentclass[11pt]{article}
\pdfoutput=1
\usepackage{preamble}
\usepackage[utf8]{inputenc}
\usepackage[T1]{fontenc}
\usepackage{xspace}
\usepackage[sort,compress,numbers]{natbib}

\renewcommand{\pt}{\ensuremath{p_{\mathrm{T}}}\xspace}
\newcommand{\Pg}{\ensuremath{\mathrm{g}}\xspace}
\newcommand{\PH}{\ensuremath{\mathrm{H}}\xspace}
\newcommand{\PV}{\ensuremath{\mathrm{V}}\xspace}
\newcommand{\PW}{\ensuremath{\mathrm{W}}\xspace}
\newcommand{\PZ}{\ensuremath{\mathrm{Z}}\xspace}
\newcommand{\PQ}{\ensuremath{\mathrm{q}}\xspace} 
\newcommand{\PAQ}{\ensuremath{\overline{\mathrm{q}}}\xspace} 
\newcommand{\PQb}{\ensuremath{\mathrm{b}}\xspace} 
\newcommand{\PAQb}{\ensuremath{\overline{\mathrm{b}}}\xspace}
\newcommand{\PQc}{\ensuremath{\mathrm{c}}\xspace} 
\newcommand{\PAQc}{\ensuremath{\overline{\mathrm{c}}}\xspace}
\newcommand{\bbbar}{\PQb{}\PAQb\xspace}
\newcommand{\bbbb}{\bbbar{}\bbbar\xspace}
\newcommand{\ccbar}{\PQc{}\PAQc\xspace}
\renewcommand{\qqbar}{\PQ{}\PAQ\xspace}
\newcommand{\mSD}{\ensuremath{m_\mathrm{SD}}\xspace}

\renewcommand{\kt}{\ensuremath{k_{\mathrm{T}}}\xspace}
\newcommand{\GeV}{\ensuremath{\,\text{Ge\hspace{-.08em}V}}\xspace}
\newcommand{\TeV}{\ensuremath{\,\text{Te\hspace{-.08em}V}}\xspace}

\title{\Large{Improving Di-Higgs Sensitivity at Future Colliders in Hadronic Final States with Machine Learning}}

\author[1]{Artur Apresyan}
\author[2]{Daniel Diaz}
\author[2]{Javier Duarte\footnote{jduarte@ucsd.edu} }
\author[3]{Sanmay Ganguly\footnote{sanmay@icepp.s.u-tokyo.ac.jp} }
\author[2]{Raghav Kansal}
\author[4]{Nan Lu}
\author[1]{Cristina Mantilla Suarez}
\author[5]{Samadrita Mukherjee}
\author[1]{Crist\'{i}an Pe\~{n}a}
\author[2]{Brian Sheldon}
\author[4]{Si Xie}
\affil[1]{Fermi National Accelerator Laboratory, Batavia, IL 60510, USA}
\affil[2]{University of California San Diego, Department of Physics, La Jolla, CA 92093, USA}
\affil[3]{ICEPP, University of Tokyo, 7-3-1 Hongo, Bunkyo-ku, Tokyo 113-0033}
\affil[4]{California Institute of Technology, Pasadena, CA 92116, USA}
\affil[5]{Department of Theoretical Physics, Tata Institute of Fundamental Research, Mumbai 400005, India}

\usepackage[firstpage=true]{background}
\backgroundsetup{contents={\parbox{6.5in}{\snowmass}}, scale=1,placement=top,opacity=1,color=black,position={3.25in,1.2in}}

\begin{document}

\flushbottom
\maketitle

\begin{abstract}
One of the central goals of the physics program at the future colliders is to elucidate the origin of electroweak symmetry breaking, including precision measurements of the Higgs sector.
This includes a detailed study of Higgs boson ($\PH$) pair production, which can reveal the $\PH$ self-coupling.
Since the discovery of the Higgs boson, a large campaign of measurements of the properties of the Higgs boson has begun and many new ideas have emerged during the completion of this program. 
One such idea is the use of highly boosted and merged hadronic decays of the Higgs boson ($\PH\to\bbbar$, $\PH\to\PW\PW\to\qqbar\qqbar$) with machine learning methods to improve the signal-to-background discrimination.
In this white paper, we champion the use of these modes to boost the sensitivity of future collider physics programs to Higgs boson pair production, the Higgs self-coupling, and Higgs-vector boson couplings.
We demonstrate the potential improvement possible at the Future Circular Collider in hadron mode, especially with the use of graph neural networks. 
\end{abstract}

\thispagestyle{empty}

\section{Introduction}
The standard model (SM) of particle physics foresees that the Higgs boson ($\PH$) has three different kinds of interactions at tree level: (i) the Yukawa interaction with quarks and leptons, (ii) the interaction with $\PW^{\pm}$ and $\PZ$ vector bosons and (iii) the trilinear and quartic Higgs self-interactions. 
Among these, observing the SM production of two Higgs bosons and precisely measuring the corresponding Higgs boson self-coupling $\lambda$ is a prime goal of future colliders such as the high-luminosity LHC (HL-LHC) and Future Circular Collider in hadron mode (FCC-hh) to fully understand the nature of electroweak symmetry breaking.
The LHC has already reported searches for a di-Higgs final state focusing on the gluon fusion (ggF) production process $\Pg\Pg\to \PH\PH$. 
Various Higgs decay channels $\bbbar\bbbar$~\cite{CMS-PAS-B2G-22-003,CMS:2022cpr,ATLAS:2018rnh,CMS:2018sxu,Aaboud:2016xco,Aad:2015uka}, $\bbbar\tau\tau$~\cite{ATLAS:2018uni,CMS:2017hea}, $\bbbar\gamma\gamma$~\cite{ATLAS:2021ifb,CMS:2020tkr,ATLAS:2018dpp,CMS:2018tla,Aad:2014yja}, and $\bbbar\PV\PV$~\cite{ATLAS:2018fpd,CMS:2017rpp}, as well as combinations of channels~\cite{Sirunyan:2018ayu,Aad:2019uzh,Aad:2015xja}, have been investigated by both ATLAS and CMS in a great detail.
On the other hand, di-Higgs production via vector boson fusion (VBF) processes at hadron colliders has been broadly studied in the theory literature~\cite{Dolan:2013rja,Ling:2014sne,Dolan:2015zja,Bishara:2016kjn,Arganda:2018ftn}, and only recently investigated experimentally~\cite{CMS-PAS-B2G-22-003,Aad:2020kub}.
Current projections~\cite{Dainese:2703572} achieve an expected significance of approximately $4.0\,\sigma$ from CMS and ATLAS combined for the full HL-LHC data set. 
Measurements of Higgs boson pair production face the difficulty of the small expected event yields even for the mode with the largest branching fraction ($\bbbar\bbbar$) as well as the presence of similar reconstructed QCD multijet events, which occur far more often. 
However, these projections do not include dedicated analyses of highly boosted hadronic final states, which may be especially sensitive to the SM and anomalous Higgs couplings~\cite{Kling:2016lay}. 

If the Higgs boson is highly Lorentz boosted, its hadronic decay products can be reconstructed as one single jet and the jet can be tagged using jet substructure techniques~\cite{Butterworth:2008iy, Abdesselam:2010pt, Kogler:2018hem, Larkoski:2017jix}.
Moreover, several machine learning (ML) methods have also been demonstrated to be extremely efficient in jet tagging and jet reconstruction~\cite{Kasieczka:2019dbj}. 
In the present work, we adopt ML algorithms to analyze boosted di-Higgs production in the four-bottom-quark final state at the FCC-hh, which is expected to produce hadron-hadron collisions at $\sqrt{s} = 100\TeV$ and to deliver an ultimate integrated luminosity of 30\,ab$^{-1}$.
We compare our ML-based event selection to a reference cut-based selection~\cite{L-Borgonovi} to demonstrate potential gains in sensitivity.
The rest of the white paper is organized as follows.
In Section~\ref{sec:boosted} and \ref{sec:ml}, we illustrate the potential of boosted Higgs channels based on the expected yields and introduce the ML methods.
In Section~\ref{sec:cut}, we describe the reference cut-based analysis and in Section~\ref{sec:gnn}, we explain our ML-based analysis.
Finally, we provide a summary and outlook in Section~\ref{sec:outlook}.

\subsection{Boosted Higgs}
\label{sec:boosted}
The hadronic final states of the Higgs boson are attractive because of their large branching fractions relative to other channels. 
While the $\bbbar\gamma\gamma$ ``golden channel'' has a 0.26\% branching fraction, the $\bbbar\bbbar$ and $\bbbar\PW\PW$ channels have a combined 58.8\% branching fraction, which often produce a fully hadronic final state.
At low transverse momentum ($\pt$), these final states are difficult to disentangle from the background, but at high $\pt$, the decay products merge into a single jet, which new ML methods can identify with exceptionally high accuracy.
Even with a requirement on the $\pt$ of the Higgs boson, the hadronic final states are still appealing in terms of signal acceptance.
The efficiency of the $\pt>400\GeV$ requirement on both Higgs bosons is about 4\% at the LHC.
Thus, the boosted $\bbbar\bbbar$ ($\bbbar\PW\PW$) channel with $\pt>400\GeV$ has $5.2$ times ($4.3$ times) more signal events than the ``golden'' $\bbbar\gamma\gamma$ channel at the LHC.
Given the higher center-of-mass energy of the FCC-hh, the boosted fraction would increase.

Based on our preliminary investigations and existing LHC Run 2 results, these boosted channels are competitive with the $\bbbar\gamma\gamma$ channel, which corresponds to an expected significance of 2.7 standard deviations ($\sigma$) with the full ATLAS and CMS HL-LHC data set.
As such, exploring these additional final states with new methods will be crucial to achieving the best possible sensitivity to the Higgs self-coupling.

\subsection{Machine Learning for Di-Higgs Searches}
\label{sec:ml}
Emerging ML techniques, including convolutional neural networks (CNNs) and graph neural networks (GNNs)~\cite{gnn,Battaglia:2016jem,DGCNN}, have enabled better identification of these boosted Higgs boson jets while reducing the backgrounds~\cite{Lin:2018cin,Qu:2019gqs,Moreno:2019bmu,Moreno:2019neq,Bernreuther:2020vhm,Sirunyan:2020lcu}.
CNNs treat the jet input data as either a list of particle properties or as an image. 
In the image representation case, CNNs leverage the symmetries of an image, namely translation invariance, in their structure.
Deeper CNNs are able to learn more abstract features of the input image in order to classify them correctly.
GNNs are also well-suited to these tasks because of their structure, and have enjoyed widespread success in particle physics~\cite{Shlomi:2020gdn,Duarte:2020ngm,Thais:2022iok}.
GNNs treat the jet as an unordered graph of interconnected constituents (nodes) and learn relationships between pairs of these connected nodes.
These relationships then update the features of the nodes in a \emph{message-passing}~\cite{DBLP:journals/corr/GilmerSRVD17} or \emph{edge convolution}~\cite{DGCNN} step.
Afterward, the collective updated information of the graph nodes can be used to infer properties of the graph, such as whether it constitutes a Higgs boson jet.
In this way, GNNs learn pairwise relationships among particles and use this information to predict properties of the jet.

Significantly, it has been shown that these ML methods can identify several classes of boosted jets better than previous methods.
For instance these methods have been used to search for highly boosted $\PH(\bbbar)$~\cite{Sirunyan:2020hwz} and $\PV\PH(\ccbar)$~\cite{Sirunyan:2019qia} in CMS.
Most recently, they have also been shown to enable the best sensitivity to the SM $\PH\PH$ production cross section and to the quartic $\PV\PV\PH\PH$ coupling in CMS using the LHC Run 2 data set~\cite{CMS-PAS-B2G-22-003}.
In this work, we study the impact of the use of these ML algorithms in future colliders like the FCC-hh. 

\section{Reference Cut-based Event Selection}
\label{sec:cut}
For the cut-based reference selection, we follow Refs.~\cite{L-Borgonovi,Banerjee:2018yxy}.
In particular, we study the configuration in which the Higgs boson pair recoils against one or more jets.
We use the \textsc{Delphes}-based~\cite{deFavereau:2013fsa} signal and background samples from Ref.~\cite{database}.
The signal sample of $\PH\PH$+jet is generated taking into account the full top quark mass dependence at leading order (LO) with the jet \pt greater than 200\GeV. 
Higher-order QCD corrections are accounted for with a $K$-factor
$K=1.95$ applied to the signal samples~\cite{Banerjee:2018yxy}, leading to $\sigma_{\PH\PH j} = 38\,\text{fb}$ for jet $\pt > 200\GeV$ and $\kappa_\lambda=1$.
The main background includes at least four b-jets, where the two $\bbbar$ pairs come from QCD multijet production, mainly from gluon splitting $\Pg\to\bbbar$.
The LO background cross section for jet $\pt > 200\GeV$ is given by $\sigma_{\bbbar\bbbar j} \text{ (QCD)} = 443.1\,\text{pb}$.

Jets are reconstructed with the anti-\kt~\cite{Cacciari:2008gp,Cacciari:2011ma} algorithm with a radius parameter $R=0.8$ (AK8) and $R=0.4$ (AK4).
The AK8 jets are formed from calorimeter energy clusters whereas the AK4 jets are formed from track elements.

\begin{figure}[htpbp]
	\centering
		\includegraphics[width=0.49\textwidth,clip=true,viewport = 0 0 504 504]{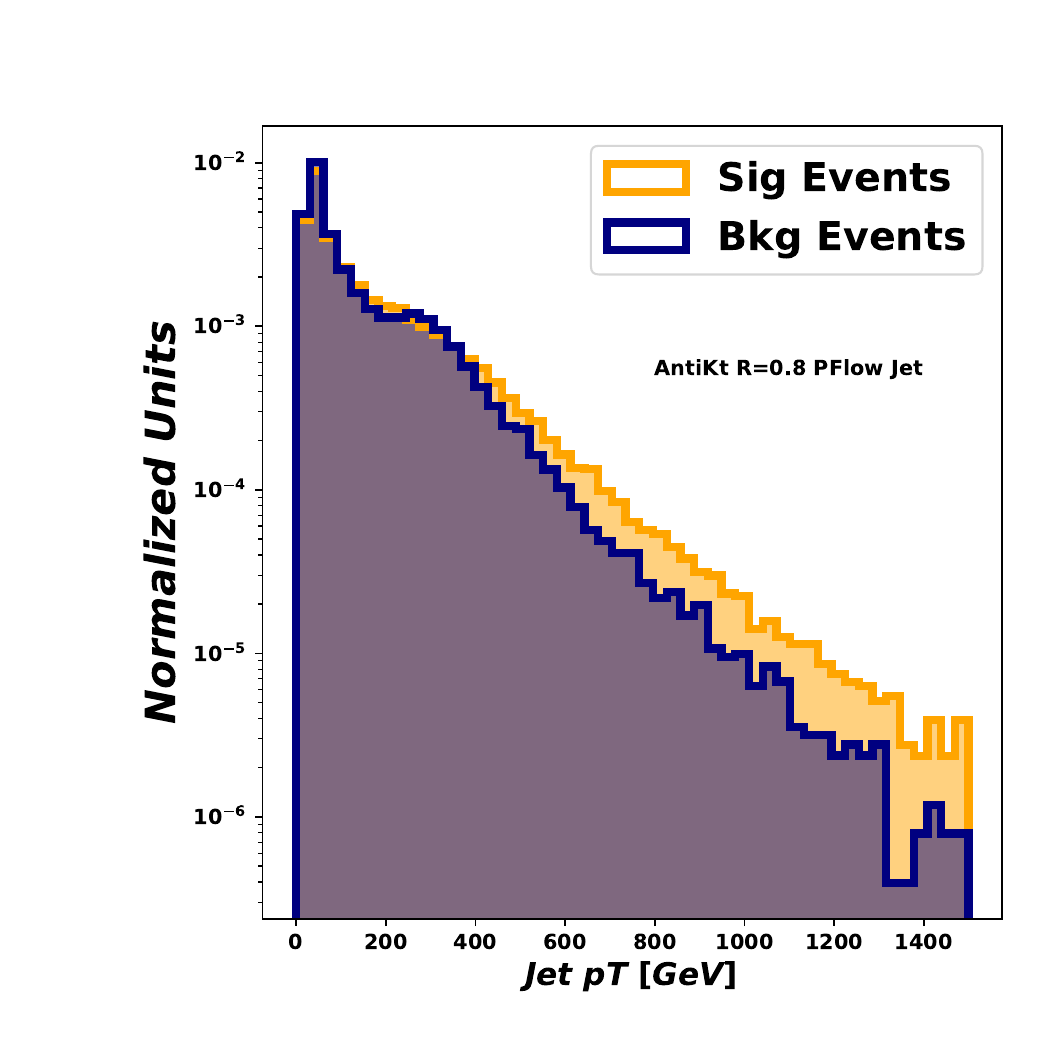}
		\includegraphics[width=0.49\textwidth,clip=true,viewport = 504 0 1008 504]{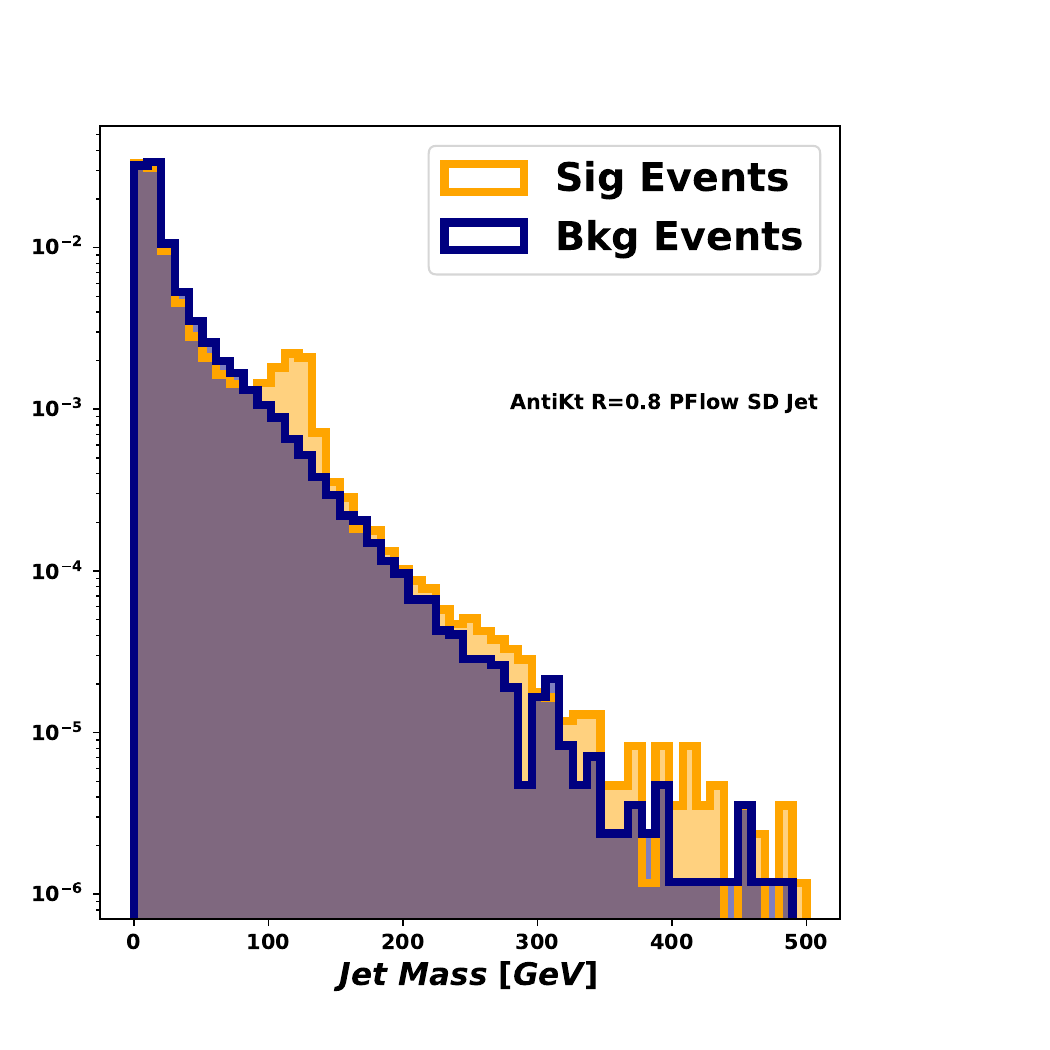}\\
		\includegraphics[width=0.49\textwidth]{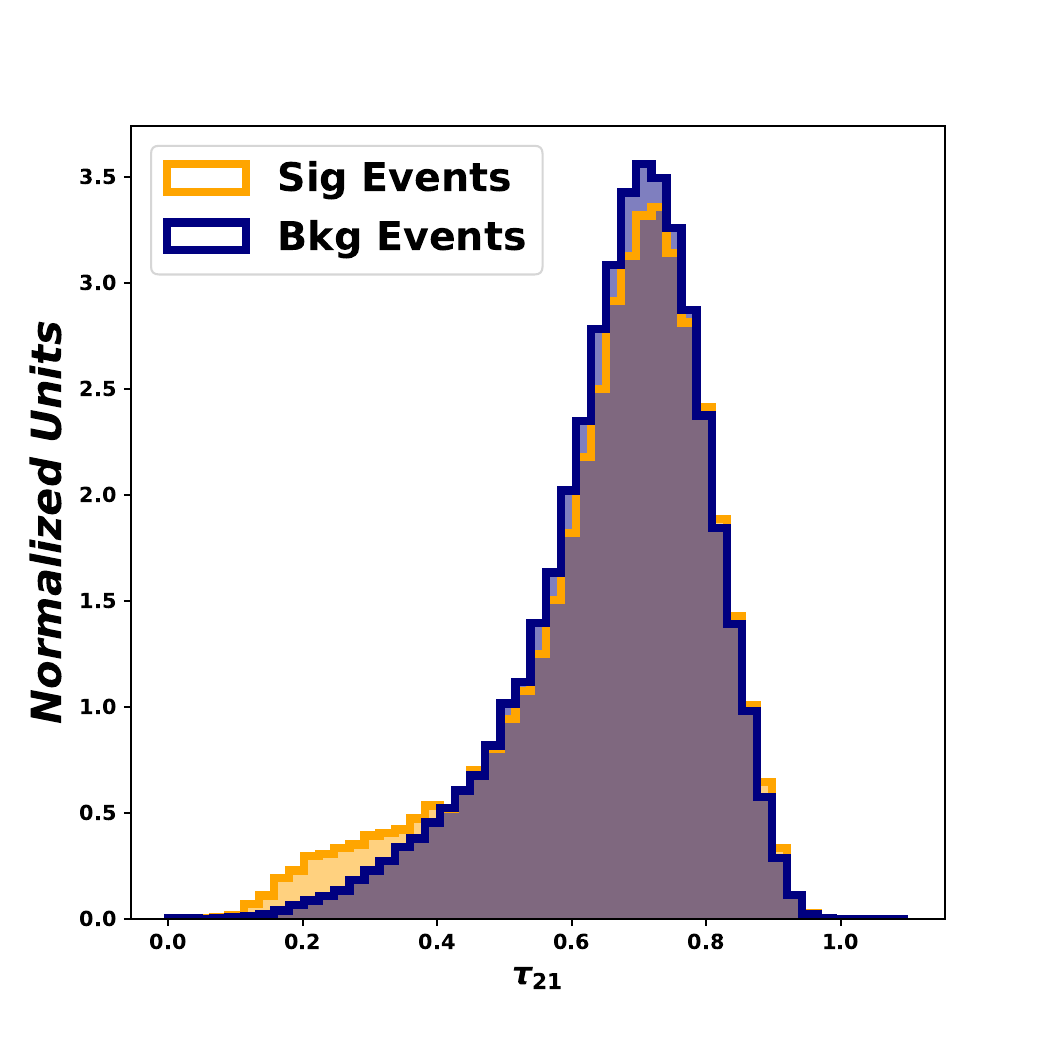}
		\caption{Jet \pt (upper left), soft-drop mass (upper right), $\tau_{21}$ (bottom) and  distribution of anti-$\kt$ $R = 0.8$ PF jets for signal and background events. 
		The shape difference plays a crucial role in identifying signal events over background.}
		\label{fig:ptmtau_dist}
\end{figure}

We require two AK8 jets with $\pt > 300\GeV$ and $|\eta| < 2.5$.
The AK8 jets are considered double b-tagged if they contain two b-tagged AK4 subjets.
This AK4 b-tagging emulation corresponds to a conservative signal efficiency of 70\%.
The two highest \pt double b-tagged AK8 jets constitute the Higgs boson candidates.
We further require the AK8 dijet system to be sufficiently boosted, $\pt^{jj} > 250\GeV$, and the leading jet to have a $\pt > 400\GeV$. 
The jet \pt and soft-drop mass \mSD~\cite{Larkoski:2014wba} distributions are shown in Figure~\ref{fig:ptmtau_dist}, along with the N-subjettiness ratio $\tau_{21} = \tau_2/\tau_1$~\cite{Thaler:2010tr}.
The two Higgs boson candidate AK8 jets are tagged by selecting jets with $\tau_{21} < 0.35$ and $100 < \mSD < 130\GeV$. 

After the selections, the expected signal ($S$) and background ($B$) yields for 30\,ab$^{-1}$ are 12\,700 and 49\,900\,000
events, yielding an approximate significance $S/\sqrt{B} = 1.8$.
The signal and background efficiencies of the cut-based selection are 1.7\% and 0.53\%, respectively. 

\section{Graph Neural Network Event Selection}
\label{sec:gnn}
We build a boosted $\PH\PH\to\bbbb$ event classifier using a GNN based on the features of all the AK4 and AK8 jet constituents (tracks and calorimeter clusters, respectively) in the event, as well as additional jet features.
We stress that this approach, an event-level classifier using the information provided in the FCC-hh samples, is conservative as we expect the largest gains in signal-to-background discrimination to arise from including lower-level detector information, including tracking and vertexing information.
Nonetheless, we can still compare this approach with a cut-based selection with access to similar information.

To define the input graph data structure or \emph{point cloud}, each of the jet constituents is treated as a node with its associated pseudorapidity ($\eta$) and azimuthal angle ($\phi$) as coordinates.
The event can then be thought of as a two-dimensional point cloud.
A graph is then formed using the k-nearest neighbor (kNN) algorithm in the $\eta$-$\phi$ plane. 
Each node has four features, namely the four components of the energy-momentum Lorentz vector.
We augment this node representation with three additional variables related to the jet as a whole.
In particular, we include the two- and one-subjettiness ($\tau_2$ and $\tau_1$) as well as $\PQb$-tagging probability. 
Hence this construction associates a feature vector of size 7 to each node. 

For the GNN architecture operation, we use the dynamic edge convolution.
The original idea was proposed for shape classification~\cite{DGCNN}, and was also used for jet classification~\cite{Qu:2019gqs}.
The message-passing (MP) operation, referred to as  
\emph{EdgeConv}, from layer $\ell$ to layer $\ell+1$ consists of the following operations
\begin{align}
 x_{i}^{\ell+1} &= \max_{j \in \mathcal{N}(i)} \left(  \Theta_{x} ( x_{j}^\ell - x_{i}^\ell ) \right) + \Phi_{x} ( x_{i}^\ell ), \\
 e_{i}^{\ell+1} &= \frac{1}{|\mathcal{N}(i)|}\left(\sum_{j \in \mathcal{N}(i)} \Theta_{e} ( e_{j}^\ell - e_{i}^\ell )  \right)  + \Phi_{e}  (e_{i}^\ell),
  \label{eqn:GraphConv}
\end{align}
where $\mathcal{N}(i)$ is the neighborhood of objects connected to object $i$, $|\mathcal{N}(i)|$ is the number of neighboring objects, $x_{i}^\ell$ are the features of node $i$ at layer $\ell$, and $e_{i}^\ell$ are the features of edge $i$ at layer $\ell$.

The implemented model has four such MP layers. 
The output dimensions of the $x$ coordinate after each layers are 3, 5, 4, and 2, respectively, whereas the dimensions of the variable $e$ are chosen to be 4, 5, 6, and 8, respectively.
The energy outputs of each layers are concatenated and passed though a MLP block to predict the output probability of the given event. 
The model is trained using a binary crossentropy loss for the classification task.
The signal events correspond to simulated $\PH\PH(\bbbb)$ events, while background events are from simulated QCD multijet production with four bottom quarks.
The optimizer used is Adam~\cite{DBLP:journals/corr/KingmaB14} with a fixed learning rate of $10^{-3}$. 
For training purposes we have used 50\,k events for training data and 10\,k events for validation data with batch size of five.

\begin{figure}[!htb]
	\begin{center}	
		\includegraphics[width=0.75\textwidth]{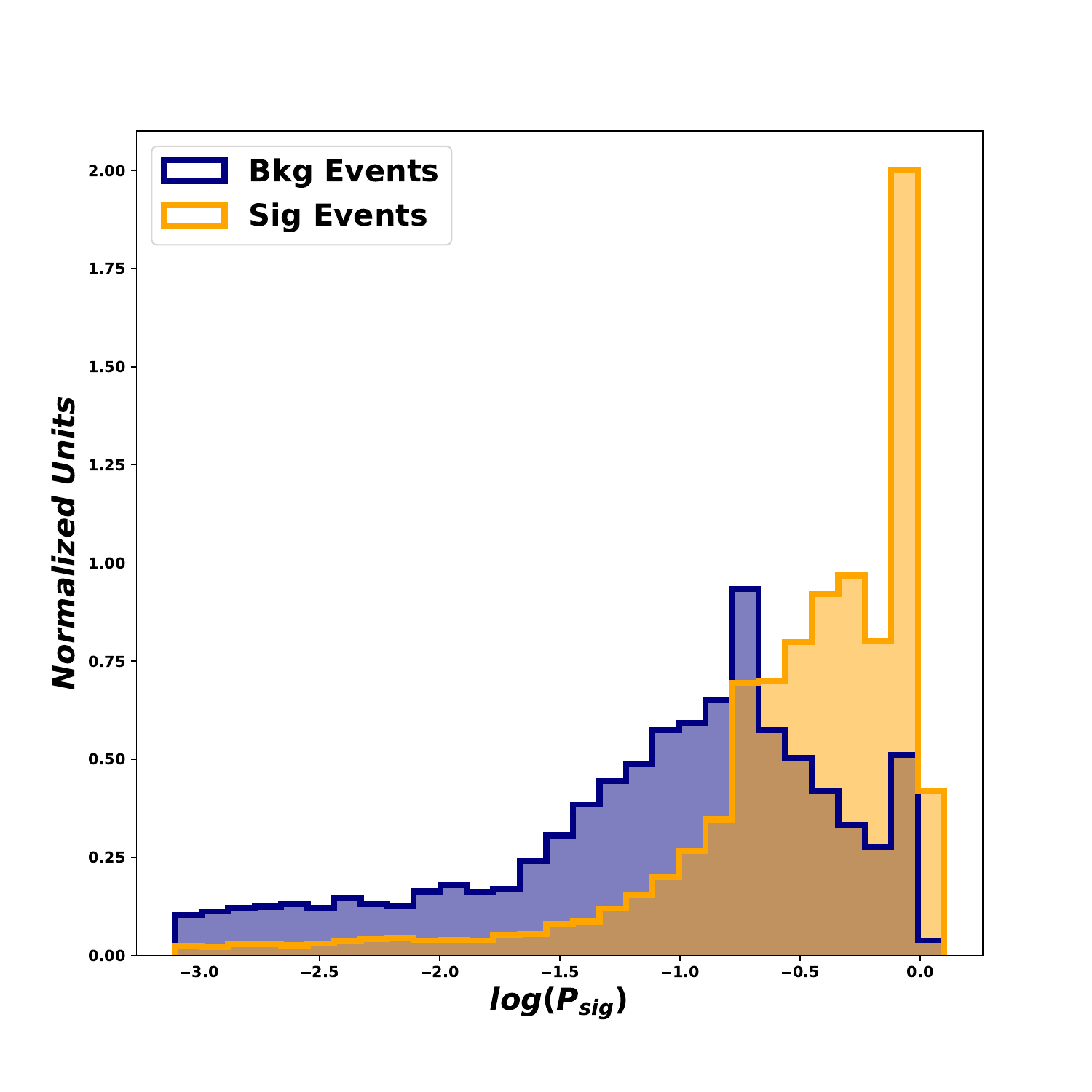}
		\caption{The distribution of natural logarithm of the events of being signal like, evaluated on the signal and background samples, respectively.}
		\label{fig:score_val}
	\end{center}
\end{figure}

\begin{figure}[!htb]
	\begin{center}	
		\includegraphics[width=0.75\textwidth]{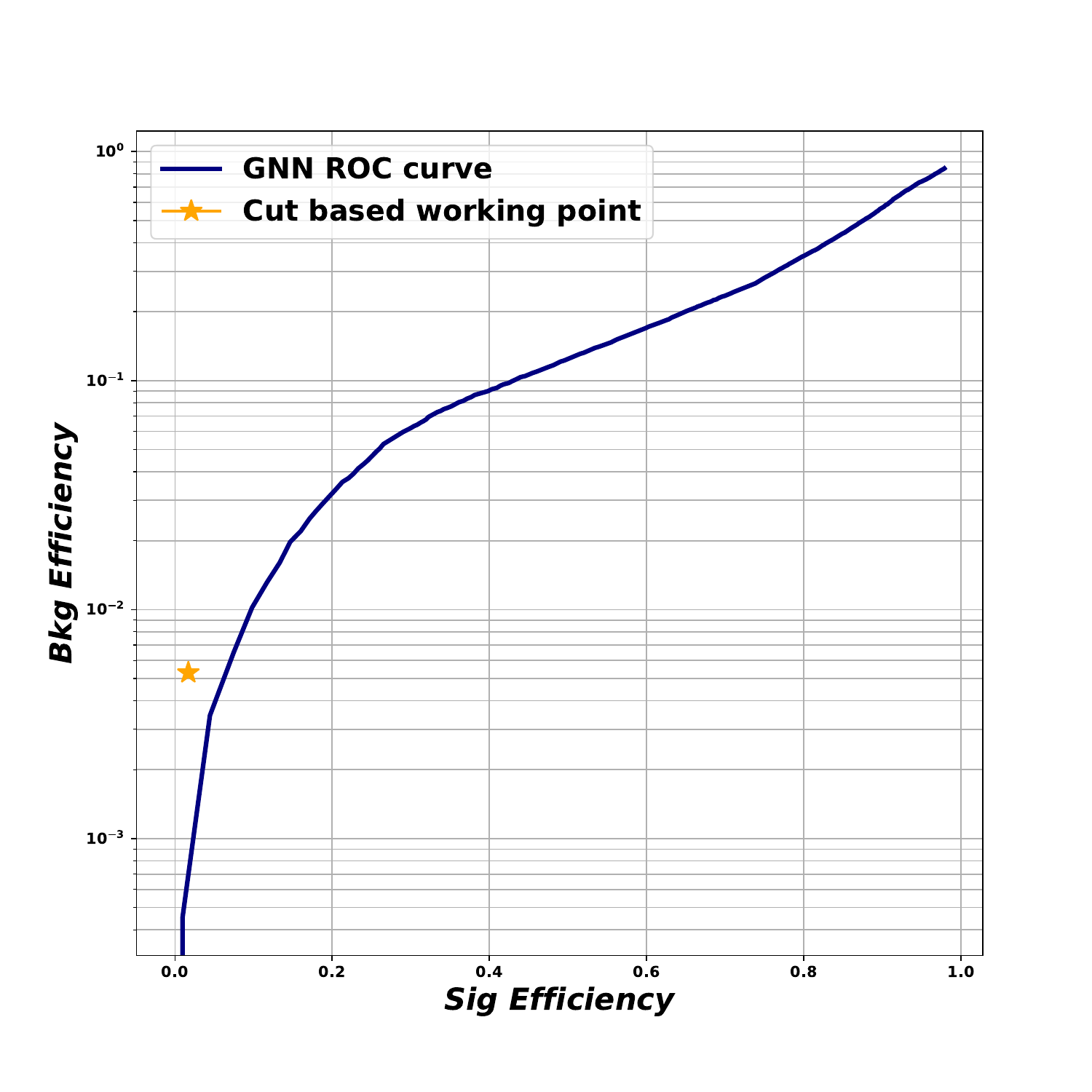}
		\caption{The ROC curve showing the ability of the NN to discriminate between signal and background. 
		With the preliminary study we achieve a signal efficiency of around 40\% of signal efficiency for a background efficiency of 9\%.}
		\label{fig:roc_curve}
	\end{center}
\end{figure}

The output of the trained network is evaluated on an independent test sample of signal and background and the logarithm of the signal-like event probability is shown in Figure~\ref{fig:score_val}. 
The distribution demonstrates that a trained network can separate the signal from background.
The level of discrimination is quantified by the receiver operating characteristic (ROC) curve shown in Figure~\ref{fig:roc_curve}. 
This preliminary training can identify signal with 40\% efficiency at the background efficiency level of 9\%.
Compared to the cut-based selection, the event-level GNN can identify $\PH\PH(\bbbb)$ signal events with an efficiency of about 6.1\% for the same background efficiency of 0.53\%, corresponding to a factor of 3 improvement.

\section{Summary and Outlook}
\label{sec:outlook}

In summary, we have investigated the feasibility of observing the production of a pair of boosted Higgs bosons in hadronic final states at the Future Circular Collider in hadron mode (FCC-hh) and improving the sensitivity with machine learning (ML) techniques.
The data sets simulated with \textsc{Delphes} corresponds to a center-of-mass energy $\sqrt{s}=100\TeV$ and an integrated luminosity of 30\,ab$^{-1}$
We focused on the four-bottom-quark final state, in which the each $\bbbar$ pair is reconstructed as a large-radius jet. 
We have studied the sensitivity using a traditional cut-based analysis as well as a selection based on an event classifier built using a graph neural network (GNN). modern GNN techniques. 
For the cut-based analysis, we leveraged the jet kinematics, substructure variables, and b-tagging for the two leading jets in the event.
For the GNN, we used lower level information, such as the jet constituents' four-momenta, as well as high-level jet substructure and b-tagging variables.
We established that a better sensitivity by a factor of 3 is achievable using the GNN as shown in Fig.~\ref{fig:roc_curve}.

Higgs boson pair production is a crucial process to characterize and measure precisely at future colliders.
In order to do so with the best precision possible, it is important to exploit all possible production and decay modes.
This includes the high-$\pt$ hadronic final states, such as $\bbbar\bbbar$, $\PW(\qqbar)\PW(\qqbar)\bbbar$, $\PW(\qqbar)\PW(\ell\nu)\bbbar$, and $\bbbar\gamma\gamma$, whose sensitivity can be improved with ML methods. 
Beyond $\PH$ jet classification, particle reconstruction~\cite{DiBello:2020bas,Pata:2021oez,Pata:2022wam}, and jet reconstruction~\cite{Guo:2020vvt}, and jet mass regression~\cite{CMS-DP-2021-017} algorithms can also be improved with ML.

Fully quantifying the impact of ML for these final states on the ultimate sensitivity achievable for the $\PH\PH$ cross section, \PH self-coupling, trilinear $\PV\PV\PH$ coupling, and quartic $\PV\PV\PH\PH$ coupling are important goals of future work.
Another important future deliverable is to consider how these ML methods may impact optimal detector design.
In this context, explainable AI methods~\cite{Mokhtar:2021bkf} can be developed to understand the physics learned by the networks, and fully exploit this in future detector design. 
Future work can also explore the impact of using symmetry-equivariant networks~\cite{Bogatskiy:2020tje,Bogatskiy:2022hub} for Higgs boson property measurements at future colliders.

\clearpage
\bibliographystyle{JHEP}
\bibliography{sample}

\end{document}